\documentclass[a4paper,12pt]{article}
\usepackage{latexsym}
\usepackage{amssymb,amsmath,amsthm}
\font\gross=cmr12 scaled \magstep3
\font\mittel=cmr12 scaled \magstep1

\font\Mittel=cmbx12 scaled \magstep1

\newcommand{\beq}{\begin{eqnarray*}}
\newcommand{\beqn}{\begin{eqnarray}}
\newcommand{\eeq}{\end{eqnarray*}}
\newcommand{\eeqn}{\end{eqnarray}}
\newcommand{\bitem}{\begin{enumerate}}
\newcommand{\eitem}{\end{enumerate}}
\newcommand{\bmatr}{\begin{array}}
\newcommand{\ematr}{\end{array}}

\newcommand{\lbeq}[1]{\label{eq:#1}}
\newcommand{\req}[1]{\ref{eq:#1}}

\newcommand{\up}{\uparrow}

\newcommand{\ts}{\textstyle}

\newcommand{\la}{\langle}
\newcommand{\ra}{\rangle}

\newcommand{\x}{{\vec x}}

\newcommand{\F}{{\cal F}}

\newcommand{\vep}{\varepsilon}
\newcommand{\vp}{\varphi}

\newcommand{\rot}{{\rm rot}}

\begin{document}
\include{epsf}

{
\thispagestyle{empty}
\baselineskip=20pt
$$ $$
\vskip 2.9cm
\centerline{\gross Exact Diagonalization Of The Fractional Quantum Hall } 
\bigskip
\centerline{\gross Many-Body Hamiltonian In The Lowest Landau Level  }

\bigskip
\bigskip
\centerline{by}
\bigskip
\bigskip
\centerline{{\mittel Detlef Lehmann}\footnote{ e-mail: 
 lehmann@math.tu-berlin.de}}
\centerline{\mittel Technische Universit\"at Berlin}
\centerline{\mittel Fachbereich Mathematik Ma 7-2}
\centerline{\mittel Sta{\ss}e des 17. Juni 136}
\centerline{\mittel D-10623 Berlin, GERMANY}
\vskip 3.2cm
\noindent{\bf Abstract:}  For a gaussian interaction $V(x,y)=\lambda\, e^{-{x^2+y^2
 \over 2r^2}}$ with long range $r>>\ell_B$, $\ell_B$ the magnetic length, we 
rigorously prove that the eigenvalues of the finite volume Hamiltonian $H_{N,{\rm LL}}= 
 P_{\rm LL} H_N  P_{\rm LL}$, $H_N=\sum_{i=1}^N  
  \bigl(-i\hbar \nabla_{x_i}-eA(x_i)\bigr)^2+\sum_{i,j; \;i\ne j} V(x_i-x_j)$, 
 $\rot A=(0,0,B)$, and $P_{\rm LL}$ the projection onto the lowest Landau level, are 
given by the following set: Let $M$ be the number of flux quanta flowing through 
the sample such that $\nu=N/M$ is the filling factor. Then each eigenvalue is 
given by $E=E(n_1,\cdots,n_N)=\sum_{i,j=1;i\ne j}^N W(n_i-n_j)$. Here
$n_i\in\{1,2,\cdots,M\}$, $n_1<\cdots<n_N$ and the function $W$ is given by
$W(n)=\lambda  \sum_{j\in\mathbb Z} e^{-{1\over 2r^2}(L{n\over M}-jL)^2}$ if the
system is kept in a volume $[0,L]^2$. The eigenstates are also explicitely given.

\bigskip
\bigskip
\vfill
\eject  }
\pagenumbering{arabic}
\baselineskip=16pt

\setcounter{equation}{0}

\bigskip
\bigskip
\indent
In this paper we consider the two dimensional many electron system in finite volume  in a constant
 magnetic field $\vec B=(0,0,B)$ described by the Hamiltonian
\beqn
H_N=\sum_{i=1}^N \left({\ts {\hbar\over i}}\nabla_{i}-e A(\x_i) \right)^2+
  \sum_{i,j=1\atop i\ne j} V(\x_i-\x_j)  \lbeq{1.1}
\eeqn
We restrict to the completely spin polarized case and neglect the Zeemann energy.
 The electron-electron interaction is assumed to be a gaussian,
\beqn
 V(x,y)=\lambda\, e^{-{x^2+y^2\over 2 r^2}}  \lbeq{1.1b}
\eeqn
which is long range, $r>>\ell_B$, $\ell_B$ being the magnetic length. The only approximation we will use
 is (see (\req{1.27c},\req{1.27e}) below)
\beqn
 \ts \int ds\, ds'\, h_n(s)\, h_{n'}(s')\, e^{-{\ell_B^2\over 2r^2}(s-s')^2}\; \approx\;
   \int ds\, ds'\, h_n(s)\, h_{n'}(s') \lbeq{1.1c}
\eeqn
where $h_n(s)=c_n H_n(s)\, e^{-{s^2\over 2}}$ denotes the normalized Hermite function.
With this approximation, the Hamiltonian $P_{\rm LL}H_NP_{\rm LL}$, $P_{\rm LL}$
 being the projection onto the lowest Landau level,  can be exactly diagonalized.
 There is the following

%
%
%
%

\bigskip
\noindent{\bf Theorem:} {\it
  Let $H_N$ be the Hamiltonian (\req{1.1}) in finite volume $[0,L_x]\times
 [0,L_y]$ (with magnetic boundary conditions (\req{RBD}), see below),
 let $A(x,y)=(-By,0,0)$ and let the  interaction be  gaussian
with long range,
\beqn
   V(x,y)=\lambda\, e^{-{x^2+y^2 \over 2r^2}}\,,\hskip 0.6cm r>> \ell_B\,.
\eeqn
Let $P_{\rm LL}:{\cal F}_N\to {\cal F}_N^{\rm LL}$ be the projection onto the lowest Landau level,
where ${\cal F}_N$ is the antisymmetric $N$-particle Fock space and ${\cal F}_N^{\rm LL}$
is the antisymmetric Fock space spanned by the eigenfunctions of the lowest Landau level. Then,
with the approximation (\req{1.1c}), the Hamiltonian $H_{N,{\rm LL}}=P_{\rm LL}H_NP_{\rm LL}$
becomes exactly diagonalizable. Let $M$ be the number of flux quanta flowing through $[0,L_x]\times
 [0,L_y]$ such that $\nu=N/M$ is the filling factor. Then the eigenstates and eigenvalues
   are labelled by $N$-tupels
 $(n_1,\cdots,n_N)$, $n_1<\cdots < n_N$ and $n_i\in \{1,2,\cdots,M\}$ for all $i$,
\beqn
 H_{N,{\rm LL}} \Psi_{n_1\cdots n_N}=(\vep_0N+E_{n_1\cdots n_N})\Psi_{n_1\cdots n_N}
\eeqn
where $\vep_0=\hbar e B/(2m)$ and
\beqn
   E_{n_1\cdots n_N}=\sum_{i,j=1\atop i\ne j}^N W(n_i-n_j)\,,
 \hskip 0.8cm W(n)=  \lambda \sum_{j\in \mathbb Z}
     e^{-{1\over 2r^2}\left(L_x{n\over M}-jL_x\right)^2}  \lbeq{E}
\eeqn
and the normalized eigenstates are given by
  $\Psi_{n_1\cdots n_N}=\phi_{n_1}\wedge \cdots \wedge \phi_{n_N}$ where
\beqn
\phi_n(x,y) &=& {\ts   {\pi^{-{1\over4}}\over  \sqrt{\ell_B L_y} }} \sum_{s=-\infty}^\infty
  e^{-{1\over 2\ell_B^2}\left(x- {n\over M}L_x-sL_x \right)^2 }
  e^{i { ( x -{n\over M} L_x -s L_x ) y / \ell_B^2 }}    \\
  &=&  {\ts   { 1\over  \sqrt{M} }  { \pi^{-{1\over4}} \over \sqrt{\ell_B L_x} } }\,
  \sum_{r=-\infty}^\infty  e^{-{1\over 2\ell_B^2}\left( y-{r\over M}L_y\right)^2}
   e^{ i    \left(x- {n\over M}L_x \right){r\over M}L_y  /\ell_B^2}
\eeqn
  }

\goodbreak

\bigskip
\noindent{\bf Proof: } We proceed in four steps: Review of the unperturbed single body
 problem, projection onto the lowest Landau level using fermionic annihilation and creation
operators, implementation of the approximation (\req{1.1c}) and finally diagonalization.

\smallskip
\noindent{\bf (i) Single Body Eigenfunctions: }

\par
We compute in finite volume with a rectangular geometry $[0,L_x]\times [0,L_y]$ and with the
Landau gauge
\beqn
   A(x,y)=(-By,0,0)  \lbeq{1.3}
\eeqn
This setup also has been used in the paper of Koma \cite{Koma}.
 The eigenfunctions $\vp=\vp_{n,k}$
   of the unperturbed single body hamiltonian
\beqn
 H_0=  \left({\ts {\hbar\over i}}\nabla-e A(\x) \right)^2 \lbeq{1.4}
\eeqn
with magnetic boundary conditions
\beqn
 \vp(x+L_x,y)=
 \vp(x,y)\,, \hskip 0.8cm \vp(x,y+L_y)=e^{ixL_y/\ell_B^2}\vp(x,y),  \lbeq{RBD}
\eeqn
  see \cite{Koma},
 are labelled by quantum numbers $n=0,1,2,\cdots$, the Landau level index, and momenta
\beqn
\ts  k={2\pi\over L_x}m,\;\;\;\;\; m=0,1,2,\cdots,M-1  \lbeq{1.5}
\eeqn
Here $M$ is the number of flux quanta flowing through the sample. Because of flux quantization,
this has to be a natural number,
\beqn
  M={L_x L_y\over 2\pi \ell_B^2}={L_xL_yB\over h/e}={\Phi\over \phi_0} \in \mathbb N   \lbeq{1.6}
\eeqn
The magnetic length is given by
\beqn
  \ell_B^2= {\hbar \over eB}  \lbeq{1.7}
\eeqn
and a flux quantum is given by
\beqn
 \phi_0=2\pi {\hbar\over e}=2,07\cdot 10^{-11} {\rm T\;cm}^2  \lbeq{1.8}
\eeqn
The energy eigenvalues are given by
\beqn
 H_0\vp_{n,k}=\vep_n\,\vp_{n,k}\,,\hskip 0.5cm  \vep_n=\hbar \ts {eB\over m}\left(n+{1\over2}\right)
      \lbeq{1.9}
\eeqn
and have an $M$-fold degeneracy. The fraction
\beqn
 \nu:= {N\over M} \lbeq{1.10}
\eeqn
is  the filling factor of the system, if $N$ denotes the number of electrons.  The explicit form of
the normalized finite volume eigenstates in asymmetric gauge is \cite{Koma}
\beqn
\vp_{n,k}(x,y)={\ts {1\over \sqrt{L_x\ell_B}}} \sum_{j=-\infty}^\infty e^{i(k+jK)x} \,h_{n,k}(y-jL_y)
   \lbeq{1.11}
\eeqn
where $K:=L_y/\ell_B^2$  and
\beqn
 h_{n,k}(y)=h_n\bigl( (y-y_k)/\ell_B\bigr)  \lbeq{1.12}
\eeqn
$y_k=\ell_B^2 k$ and $h_n$ is the normalized Hermite function,
\beqn
 h_n(y)=\ts  c_n \, H_n(y)\, e^{-y^2\over 2} \,,\hskip 0.5cm
  c_n= {1\over {^4\!\sqrt{\pi}}}{1\over \sqrt{2^n n!}} \lbeq{1.13}
\eeqn
%
%
%
%

\bigskip
\noindent{\bf (ii) Projection onto the Lowest Landau Level}

\medskip
To project $H_N$ onto the lowest Landau level, we rewrite $H_N$ in terms of fermionic
annihilation and creation operators
\beqn
H_N&=&\Bigl\{\int d^2x\, \psi^+(\x) \left({\ts {\hbar\over i}}\nabla-e A(\x) \right)^2\psi(\x)
   \nonumber\\
 &&\phantom{mm} +
  \int d^2x \,d^2x' \psi^+(\x)\psi^+(\x\,') V(\x-\x\,')\psi(\x\,') \psi(x)\Bigr\}\Bigr|_{\F_N}   \lbeq{1.2}
\eeqn
where $\F_N$ is the antisymmetric $N$-particle Fock-space. We consider the completely spin
polarized case in which only one spin direction (say $\psi=\psi_\up$) contributes and we neglect
the Zeeman energy. Introducing $a_{n,k}$, $a_{n,k}^+$ according to
\beqn
\psi(\x)=\sum_{n,k} \vp_{n,k}(\x) a_{n,k}\,,\hskip 0.5cm \psi^+(\x)=\sum_{n,k} \bar\vp_{n,k}(\x) a_{n,k}^+
  \lbeq{1.14}\\
\ts a_{n,k}=\int d^2x\, \bar\vp_{n,k}(\x)\psi(\x) \,,\hskip 0.5cm
    a_{n,k}^+=\int d^2x\, \vp_{n,k}(\x)\psi^+(\x) \lbeq{1.15}
\eeqn
 the $a_{n,k}$ obey the canonical anticommutation relations
\beqn
 \{ a_{n,k},a_{n',k'}^+\}=\delta_{n,n'}\delta_{k,k'} \lbeq{1.16}
\eeqn
and (\req{1.2}) becomes, if $H=\oplus_N H_N$,
\beqn
H&=&H_{\rm kin}+H_{\rm int} \lbeq{1.16b}
\eeqn
where
\beqn
H_{\rm kin}= \sum_{n,k} \vep_n\, a_{n,k}^+ a_{n,k} \lbeq{1.16c}
\eeqn
The interacting part becomes
\beqn
 H_{\rm int}&=& \sum_{n,k\atop n',k'} \ts \int d^2x\, d^2x'  \psi^+(\x)\psi^+(\x\,')\bar\vp_{n,k}(\x)\,
   \la n k|V|\overline{n'k'}\ra\,
    \vp_{n',k'}(\x\,')  \psi(\x\,')\psi(\x)  \phantom{mm}  \lbeq{1.21}  \\
&=&\sum_{n,k\atop n',k'}\sum_{n_1,\cdots, n_4\atop l_1,\cdots,l_4}
 (\overline{n_1 l_1};n_2l_2;\overline{nk})\,  \la n k|V|\overline{n'k'}\ra\,
    \,(n'k';\overline{n_3 l_3};n_4l_4)\, a_{n_1,l_1}^+ a_{n_3,l_3}^+     a_{n_4,l_4} a_{n_2,l_2}
 \nonumber
\eeqn
 where we used the notation
\beqn
\la n k|V|\overline{n' k'}\ra&:=&\ts \int d^2x\int d^2 x' \vp_{n,k}(\x)\, V(\x-\x\,')\, \bar\vp_{n',k'}(\x\,')
    \lbeq{1.19b} \\
(\overline{n_1 l_1};n_2l_2;\overline{n k})&:=& \ts \int d^2x \,\bar\vp_{n_1,l_1}(\x)\,\vp_{n_2,l_2}(\x)\,
    \bar\vp_{n,k}(\x)  \lbeq{1.20}
\eeqn
Now we consider systems with fillings
\beqn
 \nu\;=\;{N\over M}\;<\;1 \lbeq{1.22}
\eeqn
and restrict the electrons to the lowest Landau level.  Since the kinetic energy is constant,
we  consider only  the interacting part,
\beqn
 H_{\rm LL}&:=&P_{\rm LL} H_{\rm int} P_{\rm LL} \nonumber \\
 && \nonumber \\
  &=&\sum_{n,k\atop n',k'}\sum_{ l_1,\cdots,l_4}
 (\overline{0 l_1};0l_2;\overline{n k})\,  \la n k|V|\overline{n' k'}\ra\,
    \,(n' k';\overline{0 l_3};0 l_4)\, a_{l_1}^+ a_{l_3}^+ a_{l_4} a_{l_2} \lbeq{1.24}
\eeqn
where we abbreviated
\beqn
 a_l:=a_{0,l}\,,\hskip 0.5cm a_l^+:= a_{0,l}^+ \lbeq{1.25}
\eeqn
%
%
%
%

\medskip
\noindent{\bf (iii) The Approximation}

\par
The matrix element $\la n k|V|\overline{n' k'}\ra$ is computed in the appendix. For a
 gaussian interaction (\req{1.1b}) the exact result is
\beqn
 \la n,k|V|\overline{n',k'}\ra= \sqrt{2\pi}\ell_B\, \delta_{k,k'} \,\lambda \,r \,
  [e^{-{r^2\over2}k^2}]_M \ts
   \int  ds \int ds'   h_n(s)\, h_{n'}(s')\, e^{-{\ell_B^2\over 2 r^2}(s-s')^2}  \lbeq{1.27c}
\eeqn
where, if $k=2\pi m/L_x$,
\beqn
   [e^{-{r^2\over2}k^2}]_M:= \sum_{j=-\infty}^\infty   e^{-{r^2\over 2}(k-jK)^2 }
 =\sum_{j=-\infty}^\infty   e^{-{r^2\over 2}\left[{2\pi\over L_x}(m-jM)\right]^2 } \lbeq{1.27d}
\eeqn
is an $M$-periodic function (as a function of $m$). For a long range interaction $r>>\ell_B$, we may
approximate this by
\beqn
 \la n,k|V|\overline{n',k'}\ra &\approx& \sqrt{2\pi}\ell_B\, \delta_{k,k'} \,\lambda \,r \,
  [e^{-{r^2\over2}k^2}]_M \ts
   \int  ds \, h_n(s) \int ds' \, h_{n'}(s') \nonumber \\
 & =:&  \delta_{k,k'} \, v_k \ts
   \int ds\,  h_n(s) \int ds'\, h_{n'}(s')   \lbeq{1.27e}
\eeqn
  Then $H_{\rm LL}$ becomes
\beqn
H_{\rm LL}&=& \sum_{n,n'\atop k}\sum_{ l_1,\cdots,l_4}\ts
 (\overline{0 l_1};0l_2;\overline{n k})\,  v_k
   \int  h_n(s)ds \int  h_{n'}(s')ds'
    \,(n' k;\overline{0 l_3};0 l_4)\, a_{l_1}^+ a_{l_3}^+ a_{l_4} a_{l_2}\nonumber \\
  &=& \sum_{ k}\sum_{ l_1,\cdots,l_4}\ts
 (\overline{0 l_1};0 l_2;\overline{1_y k})\,  v_k
   \,(1_y k;\overline{0 l_3};0 l_4)\, a_{l_1}^+ a_{l_3}^+ a_{l_4} a_{l_2}
   \lbeq{1.38}
\eeqn
Here we used that
\beqn
\sum_{n=0}^\infty \ts h_n(y)\,\int  h_n(s)ds \;=\; 1 \lbeq{1.36}
\eeqn
which is a consequence of  $\sum_{n=0}^\infty h_n(y)\, h_n(s)\;=\; \delta(y-s)$.
Thus
\beqn
\lefteqn{
 \sum_{n=0}^\infty \bar\vp_{n,k}(x,y) \, \ts \int  h_n(s)ds } \nonumber \\
   &=& {\ts {1\over \sqrt{L_x\ell_B}}}\sum_{j=-\infty}^\infty e^{-i(k+jK)x}
   \sum_{n=0}^\infty  h_{n}\bigl((y-y_k-jL_y)/\ell_B\bigr)\ts \int  h_n(s)ds\nonumber \\
   &=&{\ts {1\over \sqrt{L_x\ell_B}}}\sum_{j=-\infty}^\infty e^{-i(k+jK)x} \lbeq{1.37}
\eeqn
and (\req{1.38}) follows if we define
\beqn
(\overline{0 l_1};0 l_2;\overline{1_y k}):=\int dxdy\,  \bar\vp_{0,l_1}(x,y)\, \vp_{0,l_2}(x,y)
  {\ts {1\over \sqrt{L_x\ell_B}}}\sum_{j=-\infty}^\infty e^{-i(k+jK)x}  \lbeq{1.38b}
\eeqn
These matrix elements are also computed in the appendix and the result is
\beqn
 (\overline{0 l_1};0 l_2;\overline{1_y k})=
    \delta_{m,m_2-m_1}^M {\ts {1\over \sqrt{L_x\ell_B}}} \,
   [e^{-{\ell_B^2\over 4}(l_1-l_2)^2}]_{M}  \lbeq{1.43}
\eeqn
if $k={2\pi\over L_x}m$, $l_j={2\pi\over L_x}m_j$ and $\delta^M_{m_1,m_2}=1$ iff $m_1=m_2$
  mod $M$.   In the following we write,
by a slight abuse of notation, also $\delta_{l,l'}^M$ if $l={2\pi\over L_x}m$.  Then the Hamiltonian
 (\req{1.38}) becomes
\beqn
H_{\rm LL}&=&{\ts {1\over L_x\ell_B}} \sum_{ k}\sum_{ l_1,\cdots,l_4}\ts
    \delta_{k,l_2-l_1}^M\, [e^{-{\ell_B^2\over 4}(l_1-l_2)^2}]_{M}  \,  v_k\,
   \delta_{k,l_3-l_4}^M \,[e^{-{\ell_B^2\over 4}(l_3-l_4)^2}]_{M}
   \, a_{l_1}^+ a_{l_3}^+ a_{l_4}^+ a_{l_2}\nonumber \\
  &=&{\ts {1\over L_x}} \sum_{ l_1,\cdots,l_4}\ts
   \delta_{l_2-l_1,l_3-l_4}^M \,  w_{l_2-l_1}
   \, a_{l_1}^+ a_{l_3}^+ a_{l_4} a_{l_2}  \lbeq{1.47}
\eeqn
where the interaction is given by
\beqn
 w_k:= \sqrt{2\pi}\,\lambda\,r\, [e^{-{r^2\over 2}k^2}]_M
  [e^{-{\ell_B^2\over 4}k^2}]_{M}^2  \lbeq{1.48}
\eeqn

\bigskip
\noindent{\bf (iv) Diagonalization}

\smallskip
Apparently (\req{1.47}) looks like a usual one dimensional many body Hamiltonian in momentum
space. Thus, since the kinetic energy is constant, we can easily diagonalize it by taking the
Fourier transform. For $1\le n\le M$ let
\beqn
\psi_n:={\ts {1\over \sqrt{M}}}\sum_{m=1}^M e^{2\pi i{nm\over M}}a_m, \;\;\;\;\;\;
\psi_n^+={\ts {1\over \sqrt{M}}}\sum_{m=1}^M e^{-2\pi i{nm\over M}} a_m^+ \lbeq{1.49}
\eeqn
or
\beqn
a_m={\ts {1\over \sqrt{M}}}\sum_{n=1}^M e^{-2\pi i{nm\over M}}\psi_n, \;\;\;\;\;\;
a_m^+={\ts {1\over \sqrt{M}}}\sum_{n=1}^M e^{2\pi i{nm\over M}} \psi_n^+ \lbeq{1.50}
\eeqn
Substituting this in (\req{1.47}), we get
\beqn
H_{\rm LL}&=&\sum_{n,n'} \psi_n^+ \psi_{n'}^+ \,W(n-n')\, \psi_{n'} \psi_n  \lbeq{1.51}
\eeqn
with an interaction
\beqn
 W(n)={\ts {1\over L_x}} \sum_{m=1}^M e^{2\pi i {nm\over M}} w_m  \lbeq{1.52}
\eeqn
where $w_m\equiv w_k$ is given by (\req{1.48}), $k=2\pi m /L_x$. The $N$-particle  eigenstates of
 (\req{1.51}) are labelled by $N$-tupels $(n_1,\cdots,n_N)$ where $1\le n_j\le M$ and $n_1<n_2<
  \cdots <n_N$ and are given by
\beqn
\Psi_{n_1\cdots n_N}&=&\psi_{n_1}^+\psi_{n_2}^+\cdots \psi_{n_N}^+ |{\bf 1}\ra \nonumber\\
 &=&{\ts {1\over M^{N/2}}}  \sum_{j_1\cdots j_N} e^{-{2\pi i \over M}( n_1j_1+
  \cdots n_Nj_N)} \, a_{j_1}^+ \cdots a_{j_N}^+ |{\bf 1}\ra \nonumber \\
&=&{\ts {1\over M^{N/2}}}  \sum_{j_1\cdots j_N} e^{-{2\pi i \over M}( n_1j_1+
  \cdots n_Nj_N)}\,  \vp_{0j_1}\wedge \cdots \wedge \vp_{0j_N}   \nonumber  \\
&=& \phi_{n_1}\wedge\cdots \wedge \phi_{n_N}
\eeqn
if we define
\beqn
 \phi_n(x,y):={\ts {1\over \sqrt M}} \sum_{j=1}^M e^{-2\pi i {nj\over M}} \vp_{0j}(x,y)    \lbeq{1.53}
\eeqn
The energy eigenvalues are
\beqn
 H_{\rm LL} \Psi_{n_1\cdots n_N}=E_{n_1\cdots n_N}\Psi_{n_1\cdots n_N} \lbeq{1.54}
\eeqn
where
\beqn
E_{n_1\cdots n_N}=\sum_{i,j=1\atop i\ne j}^N W(n_i-n_j)  \lbeq{1.55}
\eeqn

\bigskip
The Fourier sums in (\req{1.52}) and (\req{1.53}) can be performed with the Poisson summation
 formula. This is done in Lemma 3\hskip 0.2cm in the appendix.
 If we approximate $w_k\approx \sqrt{2\pi}\,\lambda\,r\, [e^{-{r^2\over 2}k^2}]_M$,
since by assumption $r>>\ell_B$, we  find for this $w_k$
\beqn
 W(n)&=&  \lambda \sum_{j\in \mathbb Z}
     e^{-{1\over 2r^2}\left(L_x{n\over M}-jL_x\right)^2}  \nonumber \\
 &=&  \lambda \sum_{j\in \mathbb Z}
     e^{-{1\over 2r^2}\left( \ell_B^2{2\pi n\over L_y}-jL_x\right)^2}
    \lbeq{1.63}
\eeqn
Thus  the theorem is proven $\;\blacksquare$

\goodbreak

\bigskip
\bigskip
We close with some comments. First, the approximation (\req{1.1c}) looks quite innocent.
 However, by reviewing the
computations one finds that it is actually equivalent to the approximation
 $V(x,y)=\lambda\, e^{-{x^2+y^2\over 2 r^2}}\approx \lambda\, e^{-{x^2\over 2r^2}}$. Thus
we substitute an integrable interaction by a non integrable, non symmetric one.
 As a consequence, the total energy
is no longer proportional to the volume (if the density is kept fixed) but grows as $L\times L^2$.
 Second, one may speculate that for fillings $\nu=1/q$ the ground states are labelled by
the $N$-tupels $(n_1,\cdots,n_N)$ = $\bigl(j,j+q,j+2q,\cdots, j+(N-1)q\bigr)$
 which have a $q$-fold degeneracy,
 $1\le j\le q$. A $q$-fold  degeneracy for fillings $\nu=p/q$, $p,q$ without common devisor,
   follows already from general symmetry considerations (see  \cite{Koma} or \cite{Yosh}).
 One may also conjecture that there is a gap for rational fillings, that is
\beqn
\Delta(\nu):= \lim_{N,M\to\infty\atop N/M=\nu} \bigl( E_1(N,M)-E_0(N,M) \bigr)\;
 \left\{ \begin{array}{cl} >0&{\rm if}\;\nu\in \mathbb Q \\ =0&{\rm if}\; \nu\notin \mathbb Q
   \end{array} \right.
\eeqn
Here $E_0$ is the lowest and $E_1$ the second lowest eigenvalue in finite volume.
However, we find it hard to imagine that the energy (\req{E}) distinguishes between even
and odd denominators $q$. In other words, it is questionable whether the model
 (\req{1.1},\req{1.1b}) favours the observed fractional quantum Hall fillings
 (see \cite{QHE,CF,FQH} for an overview).

\bigskip
\bigskip
\noindent{\bf Acknowledgements} It is a pleasure to thank Ruedi Seiler and Hermann
 Schulz-Baldes for several stimulating discussions on the quantum Hall effect.

%
%
%
%

\bigskip
\bigskip
\bigskip
\bigskip
\noindent{\Mittel Appendix}

\bigskip
\noindent{\bf Lemma 1:}  {\it  Let
 $ V(x,y):= \lambda\, e^{-{x^2+y^2\over2 r^2}}$.
Then one has
\beqn
 \la n k|V|\overline{n' k'}\ra= \sqrt{2\pi}\ell_B\, \delta_{k,k'} \,\lambda \,r \,
  [e^{-{r^2\over2}k^2}]_M \ts
   \int ds   \int ds' \, h_n(s)\,  h_{n'}(s')\, e^{-{\ell_B^2\over 2r^2} (s-s')^2}   \lbeq{A1}
\eeqn
where, if $k=2\pi m/L_x$,
\beqn
   [e^{-{r^2\over2}k^2}]_M:= \sum_{j=-\infty}^\infty   e^{-{r^2\over 2}(k-jK)^2 }
 =\sum_{j=-\infty}^\infty   e^{-{r^2\over 2}\left[{2\pi\over L_x}(m-jM)\right]^2 } \lbeq{1.27g}
\eeqn
is an $M$-periodic function (as a function of $m$). }

\bigskip
\noindent{\bf Proof:} We have
\beqn
\lefteqn{
\la n,k|V|\overline{n',k'}\ra=\ts \int d^2x\, d^2 x'\, \vp_{n,k}(\x)\, V(\x-\x')\, \bar\vp_{n',k'}(\x')}
  \nonumber\\
 &&  \nonumber \\
 &=&{\ts {1\over L_x\ell_B}}\sum_{j,j'} \ts \int dxdx'dydy'\, e^{i(k-jK)x-i(k'-j'K)x'}
  h_{n,k}(y-jL_y) \,h_{n',k'}(y'-j'L_y)\, V(\x-\x') \nonumber \\
 &=&{\ts {1\over L_x\ell_B}}\sum_{j,j'} \ts \int dxdx'dydy' \,e^{i(k-jK)(x-x')} e^{i(k-jK-k'+j'K)x'}
  h_n\bigl((y-y_k-jL_y)/\ell_B\bigr)\times \nonumber\\
  &&\phantom{mmmmmmm}  \,h_{n'}\bigl((y'-y_{k'}-j'L_y)/\ell_B\bigr)\,
 \lambda\, e^{-{(x-x')^2\over 2r^2}} e^{-{(y-y')^2\over 2 r^2}}  \lbeq{1.28}
\eeqn
The $x'$-integral gives  $L_x\,\delta_{m-jM,m'-j'M}=L_x\,\delta_{m,m'}\delta_{j,j'}$ if $k=2\pi m/L_x$,
 $k'=2\pi m'/L_x$, $0\le m,m'\le M-1$. Thus we get
\beqn
\la n,k|V|\overline{n',k'}\ra&=&
 {\ts { \lambda\over \ell_B}} \, \delta_{k,k'} \sum_{j} \ts \int dx \,e^{i(k-jK)x} e^{-{x^2\over 2r^2}}
 \int dydy' h_n\bigl((y-y_k-jL_y)/\ell_B\bigr) \times \nonumber \\
  &&\phantom{mmmmmmm} h_{n'}\bigl((y'-y_{k}-jL_y)/\ell_B\bigr)\,
     e^{-{(y-y')^2\over 2 r^2}} \nonumber\\
 &&  \nonumber  \\
 &=&\sqrt{2\pi} \ell_B \lambda r\, \delta_{k,k'} \sum_{j}   e^{-{r^2\over 2}(k-jK)^2 }
  \ts \int ds ds' h_n(s) \,h_{n'}(s') \,  e^{-{\ell_B^2\over 2r^2}(s-s')^2} \nonumber \\
 &=&\sqrt{2\pi} \ell_B \lambda r\, \delta_{k,k'} \, [ e^{-{r^2\over 2}k^2 }]_M
   \ts\int ds ds' h_n(s) \,h_{n'}(s') \,  e^{-{\ell_B^2\over 2r^2}(s-s')^2}   \lbeq{1.28b}
\eeqn
and the lemma follows    $\blacksquare$

\bigskip
\bigskip
\noindent{\bf Lemma 2:}  {\it  The matrix elements (\req{1.38b}) are given by
\beqn
 (\overline{0,l_1};0,l_2;\overline{1_y,k})=
  \delta_{m,m_2-m_1}^M {\ts {1\over \sqrt{L_x\ell_B}}} \,
   [e^{-{\ell_B^2\over 4}(l_1-l_2)^2}]_{M}  \lbeq{1.43b}
\eeqn
if $k={2\pi\over L_x}m$, $l_j={2\pi\over L_x}m_j$ and $\delta^M_{m_1,m_2}=1$ iff $m_1=m_2$
  mod $M$.
  }

\bigskip

\noindent{\bf Proof:} We have
\beqn
 (\overline{0,l_1};0,l_2;\overline{1_y,k})
 ={\ts {1\over \sqrt{L_x\ell_B}^3}}\sum_{j_1,j_2,j}\int dxdy\,
  e^{-i(l_1+j_1K)x}e^{i(l_2+j_2K)x} e^{-i(k+jK)x} h_{0,l_1}(y)\, h_{0,l_2}(y)\phantom{I}
    \lbeq{1.39}
\eeqn
The plane waves combine to
\beqn
\exp\left[ i\ts {2\pi\over L_x}(m_2+j_2M-m_1-j_1M-m-jM)x\right]  \lbeq{1.40}
\eeqn
and the $x$-integral gives a volume factor $L_x$ times a Kroenecker delta which is one iff
\beqn
m_2+j_2M-m_1-j_1M-m-jM=0 \lbeq{1.41}
\eeqn
or
\beqn
   m=m_2-m_1\;\;&\land&\;\; j=j_2-j_1\;\;\;\;{\rm if}\;\; m_2\ge m_1 \nonumber \\
  m=m_2-m_1+M\;\;&\land&\;\; j=j_2-j_1-1\;\;\;\;{\rm if}\;\; m_2< m_1  \lbeq{1.42}
\eeqn
Thus (\req{1.39}) becomes
\beqn
\lefteqn{
(\overline{0,l_1};0,l_2;\overline{1_y,k})=  } \nonumber \\
 &\phantom{=}& \delta_{m,m_2-m_1}^M
   {\ts  {1\over \sqrt{ {\ell_B}^3}}  {1\over \sqrt{L_x}}}\sum_{j_1,j_2}
  \int_0^{L_y} dy\, h_0\bigl( (y-y_{l_1}-j_1L_y)/\ell_B\bigr) \,
    h_0\bigl( (y-y_{l_2}-j_2L_y)/\ell_B\bigr)   \nonumber \\
  &=&\delta_{m,m_2-m_1}^M {\ts{1\over \sqrt{ {\ell_B}^3}} {1\over \sqrt{L_x}}}\sum_{j_1,j_2}
  \int_0^{L_y} dy\,  h_0\bigl( (y-y_{l_1}-j_1L_y)/\ell_B\bigr) \times \nonumber \\
 &&\phantom{\delta_{m,m_2-m_1}^M {\ts{1\over \sqrt{ {\ell_B}^3}} {1\over \sqrt{L_x}}}\sum_{j_1,j_2}
  \int_0^{L_y} dymmm}
    h_0\bigl( (y-y_{l_2}-j_1L_y+(j_1-j_2)L_y)/\ell_B\bigr) \nonumber \\
 &=&\delta_{m,m_2-m_1}^M {\ts {1\over \sqrt{ {\ell_B}^3}}{1\over \sqrt{L_x}}}\sum_{j_1,j}
  \int_0^{L_y} dy\, h_0\bigl( (y-y_{l_1}-j_1L_y)/\ell_B\bigr) \,
    h_0\bigl( (y-y_{l_2}-j_1L_y+ jL_y)/\ell_B\bigr) \nonumber \\
 &=&\delta_{m,m_2-m_1}^M {\ts{1\over \sqrt{ {\ell_B}^3}} {1\over \sqrt{L_x}}}\sum_{j}
  \int_{-\infty}^\infty dy\, h_0\bigl( (y-y_{l_1})/\ell_B\bigr) \,
    h_0\bigl( (y-y_{l_2}+ jL_y)/\ell_B\bigr) \nonumber \\
&=& \delta_{m,m_2-m_1}^M {\ts {1\over \sqrt{L_x\ell_B}}}\sum_{j}
     e^{-{1\over4\ell_B^2}(y_{l_1}-y_{l_2}+jL_y)^2}
   \nonumber \\
&=& \delta_{m,m_2-m_1}^M {\ts {1\over \sqrt{L_x\ell_B}}}\sum_{j}
  e^{-{\ell_B^2\over 4}(l_1-l_2+jK)^2} \\
&=& \delta_{m,m_2-m_1}^M {\ts {1\over \sqrt{L_x\ell_B}}} \,
   [e^{-{\ell_B^2\over 4}(l_1-l_2)^2}]_{M}  \lbeq{1.43c}
\eeqn
where $\delta_{m,m'}^M$ equals one iff $m=m'$ mod $M$ and
  equals zero otherwise $\blacksquare$

\bigskip
\bigskip
\bigskip
\noindent{\bf Lemma 3:}  {\bf (i)}  {\it    For $m\in \mathbb Z$ let
\beqn
 v_m=[e^{-{r^2\over 2}k^2}]_M=\sum_{j\in \mathbb Z}
   e^{-{r^2\over 2}({2\pi\over L_x})^2(m-jM)^2}  \lbeq{1.56}
\eeqn
and let $V(n)={1\over L_x}\sum_{m=1}^M e^{2\pi i {nm\over M}} v_m$. Then
\beqn
 V(n)={\ts {1\over \sqrt{2\pi}\, r}} \sum_{j\in \mathbb Z}
     e^{-{1\over 2r^2}\left(L_x{n\over M}-jL_x\right)^2}  \lbeq{1.57}
\eeqn
  }

\par\noindent
{\bf (ii)} {\it Let $k=2\pi m/L_x$ and let $\vp_{0,m}\equiv \vp_{0,k}$ be the single body eigenfunction
 (\req{1.11}). Then
\beqn
 {\ts {1\over \sqrt M}} \sum_{m=1}^M e^{-2\pi i{nm\over M}} \vp_{0,m}(x,y)
  &=& {\ts   {\pi^{-{1\over4}}\over  \sqrt{\ell_B L_y} }} \sum_{s=-\infty}^\infty
  e^{-{1\over 2\ell_B^2}\left(x- {n\over M}L_x-sL_x \right)^2 }
  e^{i { ( x -{n\over M} L_x -s L_x ) y \over \ell_B^2 }}   \lbeq{1.57b}  \\
  &=&  {\ts   { 1\over  \sqrt{M} }  { \pi^{-{1\over4}} \over \sqrt{\ell_B L_x} } }\,
  \sum_{r=-\infty}^\infty  e^{-{1\over 2\ell_B^2}\left( y-{r\over M}L_y\right)^2}
   e^{ i    \left(x- {n\over M}L_x \right){r\over M}L_y  /\ell_B^2}   \lbeq{1.57c}
\eeqn
 }

\bigskip
\noindent{\bf Proof:} {\bf (i)}  We have
\beqn
[e^{-{r^2\over 2}k^2}]_M&=& \sum_{j\in \mathbb Z} e^{-{r^2\over2}{M^2\over L_x^2}
   \left(2\pi {m\over M}-2\pi j\right)^2}
\eeqn
We use the following formula which is obtained from the Poisson summation theorem
\beqn
\sum_{j\in \mathbb Z}e^{-{1\over 2t}(x-2\pi j)^2} ={\ts \sqrt{t\over 2\pi}} \,
   \sum_{j\in \mathbb Z} e^{-{t\over 2}j^2} e^{ijx} \lbeq{1.58}
\eeqn
with
\beqn
\ts   x=2\pi {m\over M}\,,\hskip 0.5cm  t={L_x^2\over r^2 M^2} \lbeq{1.59}
\eeqn
Then
\beqn
 v_m&=& {\ts {1\over \sqrt{2\pi}\, r} \, {L_x\over M} } \,
   \sum_{j\in \mathbb Z} e^{-{1\over 2 r^2}{L_x^2\over M^2} j^2}
     e^{-2 \pi i{jm\over M}} \lbeq{1.60}
\eeqn
and $V(n)$ becomes
\beqn
V(n)&=&{\ts {1\over L_x}}\sum_{m=1}^M e^{2\pi i {nm\over M}}
   {\ts {1\over \sqrt{2\pi}\, r} \, {L_x\over M} } \,
   \sum_{j\in \mathbb Z} e^{-{1\over 2 r^2}{L_x^2\over M^2} j^2}
     e^{-2 \pi i{jm\over M}}   \nonumber \\
  &=& {\ts {1\over \sqrt{2\pi}\, r} \, {1\over M} } \,
  \sum_{j\in \mathbb Z} e^{-{1\over 2 r^2}{L_x^2\over M^2} j^2}
  \sum_{m=1}^M e^{2\pi i {(n-j)m\over M}}  \nonumber \\
 &=& {\ts {1\over \sqrt{2\pi}\, r} \, {1\over M} } \,
  \sum_{j\in \mathbb Z} e^{-{1\over 2 r^2}{L_x^2\over M^2} j^2}
   \, M\delta_{n,j}^M  \nonumber \\
&=& {\ts {1\over \sqrt{2\pi}\, r}  } \,
  \sum_{s\in \mathbb Z} e^{-{1\over 2 r^2}{L_x^2\over M^2} (n-sM)^2}
\eeqn
which proves part (i).

\par\noindent
{\bf (ii)}  According to (\req{1.11}) we have
\beqn
\lefteqn{
{\ts {1\over \sqrt M}} \sum_{m=1}^M e^{-2\pi i{nm\over M}} \vp_{0,m}(x,y) = }\nonumber \\
 &\phantom{=}&{\ts {1\over \sqrt M}} \sum_{m=1}^M e^{-2\pi i{nm\over M}}
  {\ts {1\over \sqrt{ L_x \ell_B}}}  \sum_{j=-\infty}^\infty e^{ i {2\pi\over L_x}(m+jM)x}
   h_{0,k} (y-jL_y) \nonumber \\
 &=&{\ts {\pi^{-{1\over4}} \over \sqrt{M L_x \ell_B}}}  \sum_{j=-\infty}^\infty \sum_{m=1}^M
   e^{ i {2\pi\over L_x}(x-{n\over M}L_x)m}  e^{i{2\pi\over L_x} jMx}
  e^{-{1\over 2\ell_B^2} \left( y-\ell_B^2 {2\pi\over L_x} m -jL_y \right)^2}\nonumber \\
 &=&{\ts   {\pi^{-{1\over4}} \over \sqrt{M L_x \ell_B}}} \sum_{m=1}^M \sum_{j=-\infty}^\infty
   e^{ i {2\pi\over L_x}(x-{n\over M}L_x)m}  e^{i{L_y\over \ell_B^2} j x}
  {\ts {1\over \sqrt{2\pi}}} \int dq\, e^{-{q^2\over 2}} e^{iq\left( {y\over \ell_B}-
       {\ell_B} {2\pi\over L_x} m -j{L_y\over \ell_B}  \right) } \nonumber \\
 &=&{\ts   {\pi^{-{1\over4}} \over \sqrt{M L_x \ell_B}}} {\ts {1\over \sqrt{2\pi}}} \int dq\, e^{-{q^2\over 2}}
   e^{iq{y\over \ell_B}}
     \sum_{m=1}^M  e^{ i {2\pi\over L_x}(x-{n\over M}L_x-q\ell_B)m} \sum_{j=-\infty}^\infty
     e^{i\left( { x\over {\ell_B}}  -q  \right){L_y\over \ell_B}  j}\,   \nonumber \\
 &=&{\ts   {\pi^{-{1\over4}} \over \sqrt{M L_x \ell_B}}} {\ts {1\over \sqrt{2\pi}}} \int dq\, e^{-{q^2\over 2}}
  e^{iq{y\over \ell_B}}
     \sum_{m=1}^M  e^{ i {2\pi\over L_x}(x-{n\over M}L_x-q\ell_B)m} \sum_{r=-\infty}^\infty
   2\pi\, \delta\left({\ts ( { x\over {\ell_B}}  -q ){L_y\over \ell_B} } -2\pi r \right)\phantom{m}
      \lbeq{1.68}
\eeqn
The delta-function forces $q$ to take values
\beqn
  q=\ts {x\over \ell_B} -2\pi r {\ell_B\over L_y}  \lbeq{1.69}
\eeqn
which gives
\beqn
\ts x-{n\over M} L_x-q\ell_B=-{n\over M} L_x+2\pi r {\ell_B^2\over L_y}={r-n\over M}\, L_x
  \lbeq{1.70}
\eeqn
Therefore the $m$-sum in (\req{1.68}) becomes
\beqn
\sum_{m=1}^M  e^{ i {2\pi\over L_x}(x-{n\over M}L_x-q\ell_B)m}=
  \sum_{m=1}^M e^{2\pi i { (r-n)m\over M}}= M\, \delta^M_{r,n} \lbeq{1.71}
\eeqn
and we get
\beqn
\lefteqn{
{\ts {1\over \sqrt M}} \sum_{m=1}^M e^{-2\pi i{nm\over M}} \vp_{0,m}(x,y) = } \nonumber \\
 &\phantom{=}&
  {\ts   {\pi^{-{1\over4}}\sqrt{M}  \over \sqrt{ L_x \ell_B}}   \sqrt{2\pi} { \ell_B\over L_y}}
     \sum_{r=-\infty}^\infty   e^{-{1\over 2\ell_B^2}\left(x-r{2\pi \ell_B^2\over L_y} \right)^2 }
  e^{i\left( {x\over \ell_B} -2\pi r {\ell_B\over L_y}\right)  {y\over \ell_B}}
     \, \delta^M_{r,n} \nonumber \\
   &=& {\ts   {\pi^{-{1\over4}}\over  \sqrt{\ell_B L_y} }} \sum_{s=-\infty}^\infty
  e^{-{1\over 2\ell_B^2}\left(x-(n+sM){2\pi \ell_B^2\over L_y} \right)^2 }
  e^{i\left( {x\over \ell_B} -2\pi (n+sM) {\ell_B\over L_y}\right)  {y\over \ell_B}} \nonumber \\
 &=& {\ts   {\pi^{-{1\over4}}\over  \sqrt{\ell_B L_y} }} \sum_{s=-\infty}^\infty
  e^{-{1\over 2\ell_B^2}\left(x- {n\over M}L_x-sL_x \right)^2 }
  e^{i { ( x -{n\over M} L_x -s L_x ) y \over \ell_B^2 }}  \lbeq{1.72}
\eeqn
 This proves (\req{1.57b}).  (\req{1.57c})  is obtained directly from (\req{1.11})
by putting $r=m+jM$.   $\blacksquare$

\end{document}